\documentclass[letterpaper, 10 pt, conference]{ieeeconf}  % Comment this line out if you need a4paper
\usepackage{amsmath}

\IEEEoverridecommandlockouts                              % This command is only needed if 
\title{\LARGE \bf
Spiking neural networks: Towards bio-inspired multimodal perception in robotics
}

\author{Katerina Maria Oikonomou\textsuperscript{1 *}, Vasiliki Balaska\textsuperscript{1}, Konstantinos A. Tsintotas\textsuperscript{1}, Christos N. Mavridis\textsuperscript{2}\\Ioannis Kansizoglou\textsuperscript{1} and Antonios Gasteratos\textsuperscript{1},
\thanks{\textsuperscript{1}K. M. Oikonomou, V. Balaska, K. A. Tsintotas, I. Kansizoglou and A. Gasteratos are with the Department of Production and Management Engineering, Laboratory of Robotics and Automation, Democritus University of Thrace, Xanthi, Greece {\tt\footnotesize emails: \{aioikono, vbalaska, ktsintot, ikansizo, agaster\}@pme.duth.gr}
}
\thanks{\textsuperscript{2}C. N. Mavridis is with the Department of Division of Decision and Control Systems, School of Electrical Engineering and Computer Science, KTH Royal Institute of Technology, Stockholm, Sweden. {\tt\footnotesize email: mavridis@kth.se}}
\thanks{This research has been co-financed by the European Health and Digital Executive Agency (HADEA), under the powers delegated by the European Commission (‘European Commission’) (project code: MASTERMINE-101091895.}
\thanks{* corresponding author}
}

\begin{document}

\maketitle
\thispagestyle{empty}
\pagestyle{empty}

%%%%%%%%%%%%%%%%%%%%%%%%%%%%%%%%%%%%%%%%%%%%%%%%%%%%%%%%%%%%%%%%%%%%%%%%%%%%%%%%
\begin{abstract}
Spiking neural networks (SNNs) have captured apparent interest over the recent years, stemming from neuroscience and reaching the field of artificial intelligence.
However, due to their nature SNNs remain far behind in achieving the exceptional performance of deep neural networks (DNNs).
As a result, many scholars are exploring ways to enhance SNNs by using learning techniques from DNNs.
While this approach has been proven to achieve considerable improvements in SNN performance, we propose another perspective: enhancing the biological plausibility of the models to leverage the advantages of SNNs fully.
Our approach aims to propose a brain-like combination of audio-visual signal processing for recognition tasks, intended to succeed in more bio-plausible human-robot interaction applications.

\end{abstract}

%%%%%%%%%%%%%%%%%%%%%%%%%%%%%%%%%%%%%%%%%%%%%%%%%%%%%%%%%%%%%%%%%%%%%%%%%%%%%%%%
\section{Introduction}
Spiking neural networks (SNNs) have been roughly studied in the last years. 
Unlike deep neural networks (DNNs), SNNs are designed to mimic the way the mammal's brain works by using spiking spikes to encode and transmit information.
Due to their nature, they are energy-efficient, making them particularly well-suited for many energy-constrained robotics applications, such as aerial, underwater and assisted living robots.
To that end, SNNs have been utilized in many robotics tasks from vision~\cite{cao2015spiking} and navigation~\cite{tang2020reinforcement} to grasping and manipulation~\cite{10092938}.
Yet, they are far behind the accuracies achieved by DNNs.
As a result, researchers have developed numerous approaches to enhance SNNs' performance.
Some of them involve transforming DNNs to SNNs~\cite{Zhang_Zhou_Zhi_Du_Chen_2019}, while others focus on refining learning techniques, thus transferring insights into learning rules from DNNs to SNNs, such as the spatiotemporal backpropagation~\cite{wu2018spatio}.
Certain approaches concentrate on the representation of input data~\cite{kundu2021hire}, while others focus on handling network parameters and effectively dealing with problematic neurons~\cite{10160749}.
Meanwhile, to further enhance robotics intelligence the integration of multiple modalities such as image and audio can be highly effective. 
In this work, we propose a novel bio-plausible approach to manage audio-visual data using different encoding schemes for each input, inspired by the way mammals synchronously process audio-visual information.

\section{Backgroung and Learning Methods}
\subsection{Multimodal Encoding Schemes}
The choice of the encoding schemes between the different modalities, \textit{i.e.}, image and audio, is crucial for leveraging the strengths of each encoding method and optimizing the performance of the SNN.
\subsubsection{Rate Coding}
Rate-coding schemes can be divided into different categories, \textit{i.e.}, count, density and population rate~\cite{auge2021survey}, with count rate being the most common one, defined by the mean firing rate as follows:
\begin{equation}
    \label{eq:rate}
    f_i^{im} = \frac{n_i}{T},
\end{equation}
where $n_i$ is the number of spikes and $T$ the time window.
In count rate encoding, the intensity of each pixel can be directly mapped to a firing rate, where higher pixel intensities result in higher firing rates. 
Rate coding, which represents information by firing rates of neurons, is well-suited for static inputs.
Images are static and spatially dense, consisting of pixel intensities. 
Hence, the intensity of each pixel can be directly mapped to a firing rate, with higher pixel intensities resulting in higher firing rates. 

\subsubsection{Time-to-First Spike (TTFS) for Audio Inputs}

Temporal encoding schemes capture the temporal dynamics of data by recording the precise time of the first spike.
Time to first spike encoding (TTFS) uses an exponential function to compute the threshold; when the input pixel exceeds it, a spike is generated.
Thus, the input pixels are translated to the exact timing of the first spike.
The threshold equation is described as:
\begin{equation}
    \label{eq:ttfs}
    f_{th}^{au} = \theta_0 \exp(-t/\tau_{th})
\end{equation}
Spectrograms are employed in audio preprocessing to convert signals into a visual representation depicting their frequency content across time.
This transformation provides a two-dimensional representation that captures the temporal dynamics of the audio data. Thus, a temporal encoding spectrogram would further enhance the temporal dynamics of the representation.
\subsection{Learning Mechanism}
The spike-timing-dependent plasticity (STDP) learning mechanism is widely used for many applications.
STDP updates the synaptic weights based on the timing difference of the pre and post-synaptic spikes.
Given the two different encoding schemes the STDP can be defined as follows:
\subsubsection{Rate based STDP}
For the rate-encoded image input, the STDP is defined as follows:
\begin{equation}
    \label{eq:stdprate}
    \Delta w_{ij}^{im}=
    \begin{cases}
    a_+ \exp(-\Delta t/\tau_+), & \text{if } \Delta t>0  \\%e^\frac{-\Delta t}{\tau_+}, & \text{if } \Delta t>0 \\
    -a_- \exp(-\Delta t/\tau_-), & \text{if } \Delta t<0.
    %e^\frac{-\Delta t}{\tau_-}, & \text{if } \Delta t<0.
    \end{cases}    
\end{equation}
where $a_+, a_-$ the parameters defining the potentiation and depression of the synaptic weights, respectively, $\tau_+, \tau_-$ the time constant parameters defining the exponential decay during potentiation and depression respectively and lastly $\Delta t = t_j^{hidden}-t_i^{im}$.
\subsubsection{Temporal based STDP}
Similarly, with the rate-encoded input, the STDP describing the temporal encoded input can be defined as follows:
\begin{equation}
    \label{eq:stdpenc}
    \Delta w_{ij}^{au}=
    \begin{cases}
    b_+(1-t_i^{hidden}/T)\exp(-\Delta t/ \tau_-), & \text{if } \Delta t>0 \\

    -b_-(1-t_i^{hidden}/T) \exp(-\Delta t/ \tau_-), & \text{if } \Delta t<0.
    \end{cases}    
\end{equation}
\section{Research Directions}
While designing SNNs, the encoding schemes and the learning mechanism need to be carefully chosen.
In this work, we propose an audio-visual SNN approach, employing different encoding schemes for each input, \textit{i.e.}, image and audio, thus aiming to enhance data representation capacities in the SNN.
To achieve that, the neurons are properly designed to handle both temporal and rate-coded inputs.
\subsection{Neuronal Dynamics}
The most common neurons used with SNNs are the leaky integrated and fire (LIF) neurons.
While deploying both image and audio inputs to the membrane potential equation, the last can take at time $t$ the following form, ensuring that both inputs influence the neurons' dynamics:
\begin{equation}
    \label{eq:mempot}
    \tau_m \frac{\Delta{V_j(t)}}{dt}= -V_j(t)+\sum_{i}w_{ij}^{im}s_i^{im}(t)+\sum_{i}w_{ij}^{au}\delta(t-t_i^{au})
\end{equation}
where $\tau_m$ is the membrane time constant, $w_{ij}^{im}$ and $w_{ij}^{au}$ are the synaptic weights of the image and audio from neurons $i$ to $j$, $s_i^{im}$ the spike trains from image and $\delta(t-t_i^{au})$ the Dirac delta function indicating the audio's contribution to the neural dynamics at $t=t_i^{au}$.
When the membrane potential of neuron $j$ reaches its threshold $V_{th}$ then it emits a spike.
Following that, the output neuron $k$ generates a spike as follows:
\begin{equation}
    \label{eq:spikegen}
    s_k^{hidden}= H(V_k(t)-V_{th}),
\end{equation}
where $hidden={\{im,au\}}$ and $H(\cdot)$ the Heaviside function.

\subsection{Combined weight update}
At time-step $t$, the combined weight update is described as follows:
\begin{equation}
    \label{eq:combw}
    \Delta w_{ij}(t) = \eta_{ij}(\Delta w_{ij}^{im}(t)+\Delta w_{ij}^{au}(t)),
\end{equation}
where:
\begin{equation}
    \label{eq:eta}
    \eta_{ij}=
    \begin{cases}
    \eta_{im}, & \text{for image input} \\
    \eta_{au}, & \text{ for audio input}.
    \end{cases}
\end{equation}
\subsection{Input Masking}
One of the main constraints of such a multimodal approach would be the choice of the proper decoding scheme, aiming to achieve high accuracies during the evaluation process.
By masking the inputs, we can compute a bias value to exploit during the evaluation process.
By masking the audio input $f_i^{au}=0$, the output of the forward pass is calculated only with the image input $output\{f_k^{im},f_k^{au}=0\}$ as well as the accuracy $a=a_{im}$, where $a$ the accuracy of the network and $a_{im}$ the accuracy of the image input.
Similarly, by masking the image input $output\{f_k^{im}=0,f_k^{au}\}$, we can compute the accuracy of the audio input $a=a_{au}$.
Following these calculations, we obtain the bias terms:
\begin{equation}
    \label{eq:bias}
    b_{im}=\frac{a_{im}}{a_{im}+a_{au}}, 
    b_{au}=\frac{a_{au}}{a_{au}+a_{im}}.
\end{equation}
Then, we can compute the biased decoding scheme for the multimodal input as follows:
\begin{equation}
    \label{eq:classif}
    C = argmax_i(b_{im}\sum_{t=0}^Ts_i^{im}(t)+b_{au}\sum_{t=0}^Ts_i^{au}(t)).
\end{equation}
\section{Conclusions}
To sum up, it is well established that by combining different modalities robots can obtain a more comprehensive understanding of their environment.
Hence, we propose a novel biologically inspired approach to handle audio-visual input with an SNN, by leveraging the advantages of different encoding schemes.
%Following the way mammals encode and handle audio-visual information, we propose an SNN that can learn how to handle different modalities effectively.
In the field of SNN research, where the focus lies on emulating various aspects of brain function, utilizing event cameras to simulate the retina and employing neuromorphic computing to develop energy-efficient platforms, it becomes crucial for us to delve deeper into how mammalian brains process different modalities.
%This understanding can enable the construction of more effective networks for deployment in robotics solutions.

\addtolength{\textheight}{-12cm}   % This command serves to balance the column lengths

%%%%%%%%%%%%%%%%%%%%%%%%%%%%%%%%%%%%%%%%%%%%%%%%%%%%%%%%%%%%%%%%%%%%%%%%%%%%%%%%

\bibliographystyle{ieeetr}
\bibliography{bibl}

\end{document}